\documentclass{article}
\usepackage[english]{babel}
\usepackage[letterpaper,top=2cm,bottom=2cm,left=3cm,right=3cm,marginparwidth=1.75cm,]{geometry}
\usepackage{amsmath}
\usepackage{graphicx}
\usepackage[colorlinks=true, allcolors=blue]{hyperref}
\usepackage{float}
\usepackage{amsthm}
\usepackage{subcaption}
\usepackage{appendix}
\usepackage[T1]{fontenc}
\usepackage{lmodern}
\usepackage{microtype}

\theoremstyle{definition}
\newtheorem{definition}{Definition}[section]

\newcounter{example}[section]
\newenvironment{example}[1][]{\refstepcounter{example}\par\medskip
   \noindent \textbf{Example~\theexample. #1} \rmfamily}{\medskip}

\title{Data valuation model for non-monetary exchanges}
\author{Julia Blyumen, Eitan Farchi}

\begin{document}
\maketitle

\begin{abstract}
In the evolving landscape of data product exchange platforms, traditional economic valuation models fall short due to the non-rival nature of data and the prevalence of non-monetary data product exchanges. This paper introduces a normative, choice-based metric for valuing data products within intracompany exchanges, where conventional pricing mechanisms are absent. By modeling consumer attention and preferences, the proposed metric quantifies the value of data offerings based solely on user selection behavior, without relying on cost, demand, or competitive pricing data. We show that this metric can be formally cast as a cooperative game with a closed-form Shapley value, providing a principled and fairness-based allocation of value across offerings. The model rewards uniqueness and discriminative consumption, effectively addressing the limitations of popularity-based metrics and incentivizing the creation of high-value, long-tail data products. Through theoretical analysis and illustrative examples, the metric is shown to align with economic principles, support equitable valuation, and contribute to a robust framework for measuring gross data product value. Future research directions include exploring bundling strategies and quantifying product complementarity.   
\end{abstract}

\section{Introduction}


\subsection{Challenges in Valuating Data Products}

To lay the groundwork for a new valuation framework, we begin by examining the multifaceted challenges inherent in valuating data products. These include the unique characteristics of data as a non-rivalrous resource, the limitations of bundling and popularity-based metrics in digital marketplaces, and the inadequacy of traditional economic models in non-monetary exchange contexts. Together, these issues underscore the need for alternative metrics that reflect user behavior, product distinctiveness, and fair recognition of product value.

\subsection{Understanding Data Products and Data Product Exchanges}
Data has become one of the most valuable digital assets for organizations. To fully realize that value, organizations have made the strategic shift to treating data as a product—purpose-built and managed for specific audiences and their use cases. Data product exchange platforms serve as the marketplace that connects data producers and consumers, enabling the discovery, access, and reuse of data products to drive business growth.\par
Data product exchanges should support the creation and tracking of metrics that reflect the potential of data products to improve business outcomes. By making the value of data products more visible, these exchanges can help both producers and consumers make informed investment decisions—encouraging participation and enabling more effective matching of supply and demand.\par
However, the industry continues to struggle with defining clear and consistent valuation metrics for data products. This challenge is not unique to data, but extends to digital products and their marketplaces more broadly.\par
While traditional economic principles such as supply, demand, cost, and price still apply, the non-rivalrous and non-scarce nature of digital products introduces unique complexities in determining their value.
\subsection{The Non-Rivalrous, Non-Scarce Nature of Data}
The non-rivalrous and non-scarce nature of digital products means that, once created, they can be consumed repeatedly at little to no additional cost to the producer, which theoretically should provide great benefits to the consumers and producers alike. Interestingly, it is this very property that creates serious pricing issues for the various digital products industries, including data products. For example, digitized music has been blamed for the decline in music sales since 2001. The availability of digital music is said to threaten the incentives for innovation and creativity itself. This is because when the cost of reproducing the products is close to zero, then close to zero prices are expected, leaving no incentives for the producers to create digital products in the first place \cite{brynjolfsson_innovation_2007}.\par
\subsection{Limitations of Bundling as a Default Strategy in Digital Marketplaces}
\label{bundling}
To address the pricing challenges of digital products, many digital marketplaces adopt bundling strategies—pricing a collection of products together rather than individually. Research shows that, by the law of large numbers, it is often easier to determine an optimal price for a bundle than for each individual product. In equilibrium, the profit-maximizing price for a large bundle tends to be low enough that most consumers interested in any part of the bundle will purchase the whole package, even if they use only a small portion of it. For instance, most PC users buy Microsoft Office, even if they don’t use all its applications, or even not all of the features of the applications that they do use. Bundling has been proven to maximize the consumers’ attention and producers’ profitability \cite{bakos1999bundling}.\par
A key challenge with the bundling approach is how to fairly distribute revenue among producers based on their individual contributions. Bundling inherently obscures information about how consumers value each product within the bundle. Without this visibility, it becomes difficult—if not impossible—to allocate profits in a way that accurately reflects each producer’s value creation. As a result, producers have little incentive to invest in improving their products, since high- and low-quality products are monetized equally within the bundle, undermining motivation for differentiation and innovation. \cite{brynjolfsson_innovation_2007}.\par

\subsection{Limitations of Popularity-Based Metrics}
\label{sectionLongTail}

Many marketplaces estimate individual contributions based solely on access counts—such as views or downloads—if one is willing to assume that the access counts signal popularity, and that popularity is a measure of value. This method is powerful in the context when all the products are approximately equal in value. However, when products vary in value, then pricing based on access counts becomes misleading \cite{jain_equilibrium_2010}. It also opens the door to manipulation, as producers can artificially inflate download counts to distort perceived value \cite{brynjolfsson_innovation_2007}.\par

\begin{figure}[htpb]
    \centering
\includegraphics[width=0.5\linewidth]{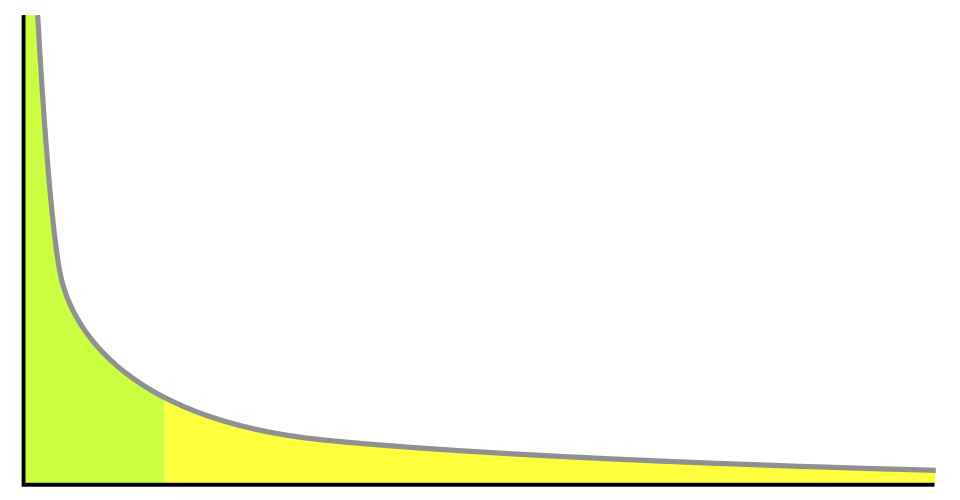}
    \caption{\label{longTail}An example of a power law graph showing popularity ranking. To the right (yellow) is the long tail; to the left (green) are the few that dominate \cite{wiki:longtail}.}
\end{figure}

On top of it, this approach penalizes the producers of less frequently used, or “long-tail,” products. The economic model behind the “long-tail” concept states that there is a small number of products used by many consumers (big head) and many products used by only a small number of consumers (long tail). When you subtotal the value of big-head products and the value of long-tail products, the two subtotals are approximately equal, each representing about half of the total value. For example, if there are 2 products that each are used by 50 consumers and 50 products that are used only by 2 consumers each, the value of each group will be half of the overall value (see Figure \ref{longTail}).\par
\par
If there are no incentives for the producers to build “long-tail” products, half of the total value is lost. Therefore, the good value metric should incentivize the production of “long-tail” products equally well as “big-head” products, which popularity metrics fail to provide.\par
It is also interesting to observe that, unlike regular products that need to be reproduced or duplicated as many times as there are consumers, digital products don’t need to be duplicated. Especially when it comes to data, enterprises strive to reduce and eliminate data duplication. Therefore, it is intuitive and reasonable that a higher value is assigned to unique data products, while less or negative value is assigned to duplicate ones. Popularity metrics incentivize producers in a different direction. If a certain song is very popular, that incentivizes the producers to create more songs like that one.

\subsection{The Case for New Metrics in Non-Monetary Exchanges}
\label{non-monetary}

While certain data product exchanges directly monetize data products by selling them, the majority are internal to the organization and provide free access to these data products for its employees, partners, and customers, unlocking the value indirectly via improvements in productivity and decision-making.\par
For these intracompany exchanges, information such as costs, demand and competitive pricing may not be available, making it very difficult or impossible to apply traditional product valuation approaches such as cost-based, value-based, investment-based or competition-based approaches.\par
These non-monetary environments require complete decoupling of valuation techniques from pricing.

\section{A Normative Model for Data Product Valuation}

In this paper, we propose a metric of data product value that relies solely on consumer choices, disregarding prior knowledge of costs, competitive pricing, and return on investment, removing barriers for businesses to start making insights-driven decisions about their data products (see \ref{non-monetary}). Our approach is based on a formal, theoretical model of consumer behavior, in contrast to statistical methods that infer preferences from empirical data \cite{Chen04}.\par
The proposed metric does not rely on aggregate packaging strategies. By valuing products individually through user choice, it enables fair attribution and thereby overcomes the known limitations of bundling (see \ref{bundling}).\par
We demonstrate how the metric exhibits the intuitively desirable characteristics described by the formal normative analysis. Specifically, we demonstrate how the metric enables data product exchanges to more accurately represent the value of infrequently used ``long-tail'' (see \ref{sectionLongTail} ) products and discourage duplication.


\subsection{Attention as a Unit of Value}
\begin{figure}[htpb]
\begin{subfigure}{0.49\textwidth}
\centering
\includegraphics[width=0.49\linewidth]{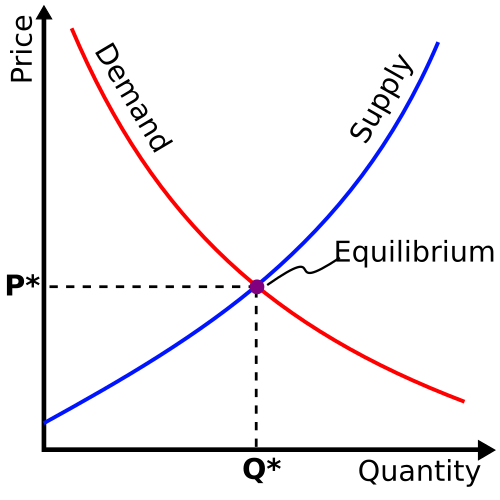} 
\caption{{\label{equilibrium}}Supply and demand curves with economic equilibrium of price and quantity  \cite{wiki:supplydemand}.}
\end{subfigure}
\begin{subfigure}{0.49\textwidth}
\centering
\includegraphics[width=0.49\linewidth]{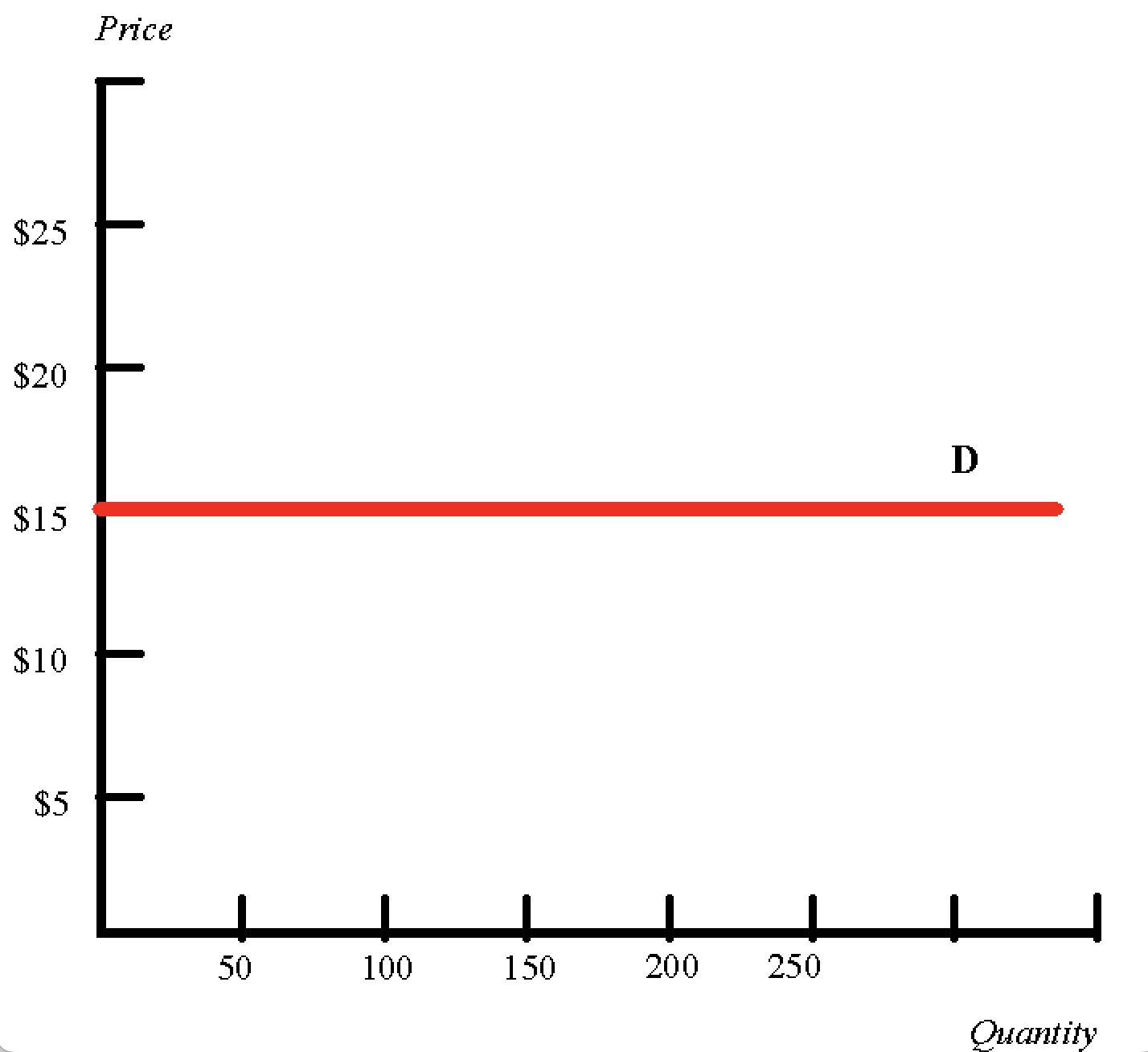}
\caption{{\label{elastocoty}}Perfect elasticity where the quantity that can be supplied is infinite \cite{wikimedia:perfectlyelastic}.}
\end{subfigure}
\end{figure}

Conventionally, valuation and pricing rely heavily on the equilibrium point of demand and supply curves (see Figure \ref{equilibrium}). However, this calculation becomes unfeasible with the near-infinite supply of digital products, such as data, which is often referred to as "almost perfectly elastic" (see Figure \ref{elastocoty}).\par
With that being said, we shall consider that demand is not infinite and is limited by consumers’ time and attention. “Capturing market share” or “capturing attention” is becoming an increasingly important consideration in digital marketplaces today and economics in general \cite{heitmayer2025second}. Therefore, valuating a data product or a bundle of products by its consumption, or more precise, the share of the attention the consumers are willing to allocate to a specific offering, is a logical strategy.\par
Consequently, we express an offering's value as the inverse of the usage of all other offerings. Here we assume that the more consumers allocate their differentiated attention to a specific offering, the more valuable the offering is. If a consumer accepts all offerings, their choice has little impact on the offering's value. The more discriminative the consumer is, the greater their choice influences the valuation of offerings.\par
Similar approaches have been explored. For instance, Brynjolfsson and Zhang suggest conducting small experiments to determine if consumers would forfeit access to most offerings for access to a few select ones \cite{brynjolfsson_innovation_2007}.\par
Our approach differs, as it incorporates attention into the offering valuation metric itself, eliminating the need for separate experiments.

\subsection{Model Setup and Key Definitions}
We consider a system with a set of $n$ data product \textbf{{offerings}} $O = \{o_1, \ldots, o_n~|~o_i \in R^k\}$,  available to a set of $m$ \textbf{{consumers}} $C = \{c_1, \ldots, c_m\}$ where each consumer can \textbf{{subscribe}} ($sb$) to one or more offerings. Our objective is to quantify consumer preferences for these offerings in order to infer the relative value of each.\par
A key indicator of consumer preference is whether a consumer $c_i$ subscribes to an offering $o_i$ or not. (It’s worth noting that this model can be generalized to other contexts by redefining the key event. For example, in a retail setting, the same abstraction could apply to a set of shops, where the key event is whether a consumer $c_i$ regularly visits a shop $o_j$.)\par
The system configuration is represented by a relation $sb : C \rightarrow P(O) $, that associates each consumer with 0 or more offerings that she subscribes to.\par
We make the following assumptions about the consumers. We assume that given two offerings $o_i$ and $o_j$ any offering $\lambda \times o_i + (1-\lambda) \times o_j, 0 < \lambda < 1$ can be created.  We make the following assumptions about the consumers. If $o_i \in sb(c_l)$ and  $o_j \in sb(c_l)$ then the consumer $c_l$ exhibits indifference between $o_i$ and $o_j$.  On the other hand, if $o_i \in sb(c_l)$ and $o_j \notin sb(c_l)$ then for any possible alternative offering $\lambda \times o_i + (1-\lambda) \times o_j, 0 < \lambda < 1$ the $c_l$ will prefer $o_i$ to $\lambda \times o_i + (1-\lambda) \times o_j, 0 < \lambda < 1$.\par
Let’s examine some normative heuristics that may explain why a consumer chooses—or does not choose—a particular offering.\par
\begin{quote}
  \textbf{Single priority}: A consumer may have a single goal (e.g., user segmentation) and subscribe to multiple demographics databases, indicating no strong preference. Conversely, choosing a single database suggests it is the best fit for the use case.\par
\textbf{Multiple priorities}: A consumer with multiple goals (e.g., user segmentation and equipment failure analysis) may subscribe to multiple offerings, implying equal prioritization of these different goals. Opting for a single offering indicates a higher value of one goal over others.\par
\textbf{Simultaneous product selection}: A consumer subscribing to two offerings—such as a user demographics database and a usage dataset—at the same time likely requires both to derive value (e.g., for affinity analysis).
\end{quote}
In all scenarios, the more uniquely an offering is chosen at the expense of others, the more valuable it is. Consequently, we propose a scoring system for offerings as an inverse function of consumer preferences for all other offerings.\par
\begin{definition}
 We define the value score of an offering as \[v(o_i) = \sum_{o_i \in sb(c_j)}  \frac{1}{|sb(c_j)|}\]    
\end{definition}


\begin{example}
Consider a scenario with two data product offerings, \( O = \{o_1, o_2\} \), and two consumers, \( C = \{c_1, c_2\} \). Consumer \( c_1 \) strictly prefers \( o_1 \) over \( o_2 \), subscribing only to \( o_1 \), i.e., \( sb(c_1) = \{o_1\} \). Consumer \( c_2 \) is indifferent between the two and subscribes to both, i.e., \( sb(c_2) = \{o_1, o_2\} \).

We compute the value score for each offering using the proposed metric:

\[
v(o_1) = \sum_{o_1 \in sb(c_j)} \frac{1}{|sb(c_j)|} = \frac{1}{1} + \frac{1}{2} = 1.5
\]

\[
v(o_2) = \sum_{o_2 \in sb(c_j)} \frac{1}{|sb(c_j)|} = \frac{1}{2} = 0.5
\]

As expected, \( o_1 \) receives a higher score, reflecting its stronger and more selective preference among consumers.
\end{example}

\subsection{Cooperative game formulation}


It is worth noting that \textbf{this model can be naturally cast as a cooperative game with the Shapley value embedded in its structure}. The Shapley value is a foundational concept in cooperative game theory that provides a principled method for fairly attributing value to individual contributors in a collective outcome~\cite{shapley1953value}. It is one of the most widely studied solution concepts in economics and offers a unique allocation based on each participant’s marginal contribution across all possible coalitions. In this setting, the structure of the game admits a simple closed-form expression for the Shapley value.
\par
While the Shapley value is well-defined for any cooperative game, its computation depends on the structure of the underlying model. We build on the cooperative game-theoretic framework introduced by Kleinberg et al.~\cite{10.5555/1028128.1028156}, adapting it to a data product setting in which consumer subscription behavior is used to infer and allocate value across offerings.



\par
\cite{10.5555/1028128.1028156} defines the \textbf{recommendation cooperative game} as a cooperative game $(N, v)$, where:
\begin{itemize}
  \item $i$ is a player and $j$ is an item; $i \in N$.
  \item Each player $i$ recommends a set of items, denoted $B_i$.
  \item The total set of unique items recommended by any player $i \in N$ is $B = \bigcup_{i \in N} B_i$.
  \item $S$ is a coalition of players $S \subseteq N$. The value of a coalition is $v(S) = \left|\bigcup_{i \in S} B_i\right|$.
  \item $\{k \mid j \in B_k\}$ represents the set of players who recommend item $j$, and $\left| \{k \mid j \in B_k\} \right|$ its size.
  \item The Shapley value of a player $i$ is
  \[
  v(i) = \sum_{j \in B_i} \frac{1}{\left| \{k \mid j \in B_k\} \right|}.
  \]
\end{itemize}
Intuitively, each item contributes one unit of value, which is shared equally among all players who recommend it; thus, the contribution of an item decreases as more players recommend it.

\par While the original formulation of the recommendation cooperative game treats consumers as players who recommend items, we reverse this perspective to better suit our goal. Offerings become the players, each associated with the set of consumers who subscribe to it, allowing us to analyze value through shared and unique audience reach. In this view, it is as if once a customer chooses an offering, the offering “recommends” that customer, preserving the cooperative structure while aligning with our product-centric setting. Intuitively, each consumer contributes one unit of value, shared equally among all offerings to which the consumer is subscribed.

\par \textbf{We now recast this as a cooperative game where players are offerings}, reusing the previously defined offerings $o_i$ and consumers $c_j$. Each offering $o_i$ is associated with the set of consumers subscribed to it, denoted $B_{o_i} = \{c_j \mid o_i \in sb(c_j)\}$.

\par Equivalently, $sb(c_j)$ denotes the set of offerings subscribed to by consumer $j$, and we have the correspondence $c_j \in B_{o_i} \Leftrightarrow o_i \in sb(c_j)$. Under this correspondence, $\left| \{k \mid c_j \in B_k \} \right| = |sb(c_j)|$, the number of offerings subscribed to by consumer $j$. Substituting into the Shapley value expression yields the original definition of the offering value score: 
\[
v(o_i) = \sum_{c_j \in B_{o_i}} \frac{1}{\left| \{k \mid c_j \in B_k \} \right|}
= \sum_{o_i \in sb(c_j)} \frac{1}{|sb(c_j)|}.
\]
\par This formulation places the model within the well-established Shapley value framework, providing a principled interpretation and a fairness-based allocation of value.

\section{Theoretical Properties and Observations}

\subsection{Incentivizing Long-Tail Data Products}
Let's examine the metric performance using a simple but realistic scenario.\par

\begin{figure}[htpb]
    \centering
  \includegraphics[width=1\linewidth]{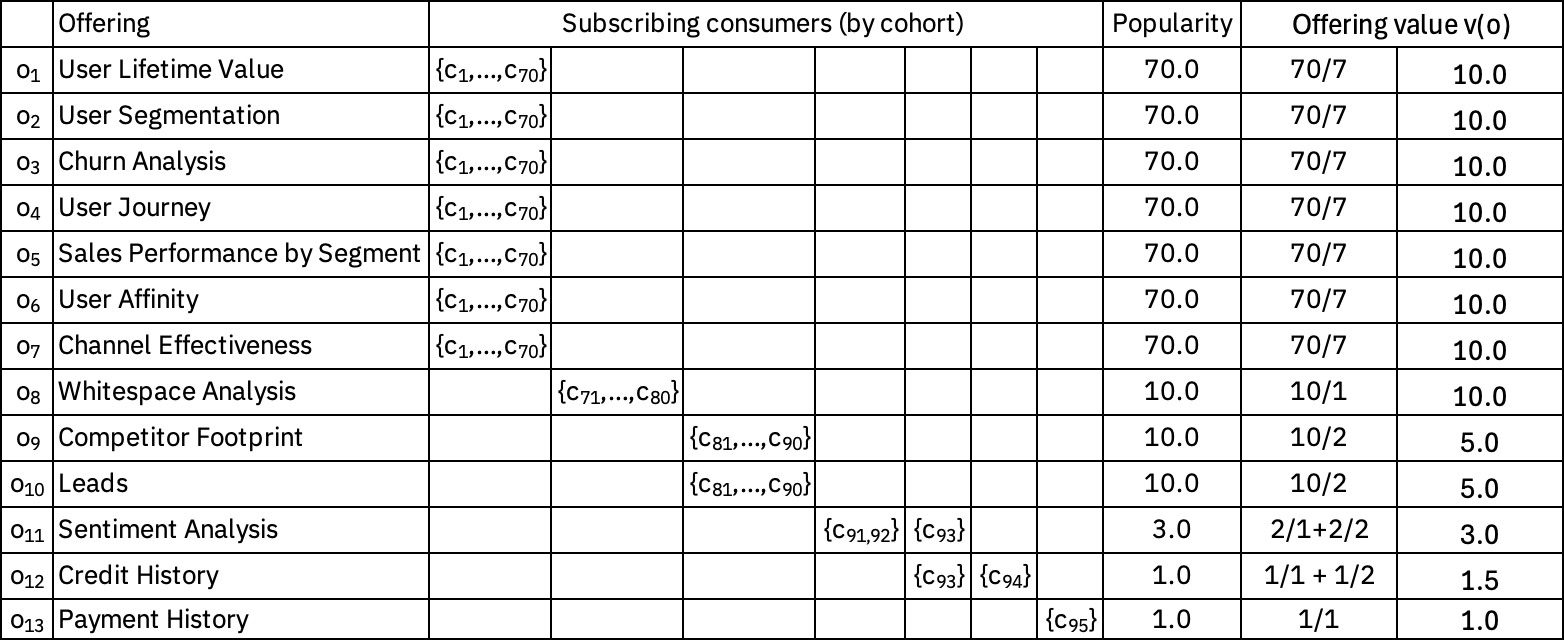}
    \caption{\label{simulation}}
\end{figure}

  We assume that there are 13 offerings and 95 consumers. The first 7 offerings ($o_1$ to $o_7$) are the most popular, subscribed to by 70 consumers ($c_1$ to $c_{70}$). The eighth offering ($o_8$) is uniquely chosen by a distinct group of 10 consumers ($c_{71}$ to $c_{80}$). Another 10 consumers ($c_{81}$ to $c_{90}$), subscribe to both $o_8$ and $o_9$, indicating partial uniqueness for $o_9$. The remaining 5 consumers  ($c_{91}$ to $c_{95}$) each choose one or two of the remaining offerings ($o_{10}$ to $o_{13}$) (see Figure \ref{simulation}).

In other words, we observe three distinct groups of offerings:

\begin{enumerate}
  \item 
Highly popular but not uniquely chosen offerings ($o_1$ to $o_7$) 
  \item 
  Moderately popular and uniquely chosen offerings ($o_8$ and $o_9$). 
  \item 
  Unpopular but uniquely chosen offerings ($o_{10}$ to $o_{13}$)
  \end{enumerate}

Based on the subscription patterns, the expected value distribution is as follows:\par

Offerings $o_1$ to $o_7$ should have relatively high value due to their popularity, but this value should be \textit{discounted} because they are not uniquely chosen. Offering $o_8$ should receive high value as it is \textit{most uniquely chosen} by a dedicated group. Offering $o_9$ is also uniquely chosen, but less so than $o_8$, so its value should be \textit{lower than that of $o_8$}. Offerings $o_{10}$ to $o_{13}$ are chosen by fewer consumers but are uniquely selected, so they should receive \textit{lower value} than $o_8$ and $o_9$.

We calculate the score of each offering using our  metric, comparing it to the simple popularity metric (a number of subscriptions).\par

\begin{figure}[htpb]
    \centering
\includegraphics[width=1\linewidth]{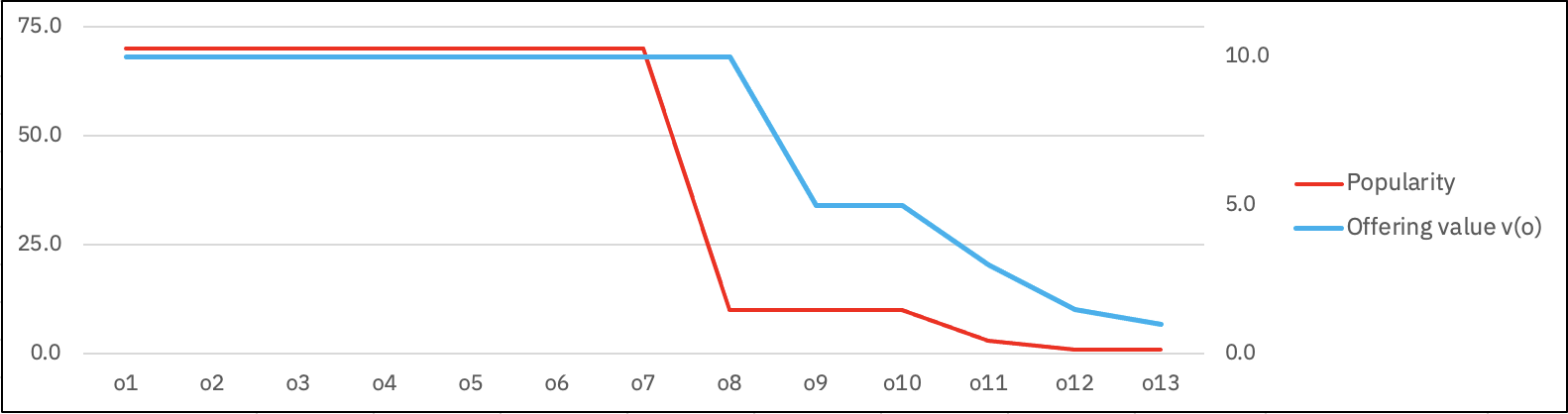}
    \caption{\label{simulation_results}}
\end{figure}

The analysis confirms that our metric supports the normative logic. It recalibrates the values, giving non-popular but uniquely chosen offerings a boost, allowing them to score comparably to popular but non-unique offerings (see Figure \ref{simulation_results}). This demonstrates that the metric performs well and aligns with the original intuitions proposed.

\subsection{Incentivizing Data Product Producers}

A key question in evaluating the effectiveness of the proposed metric is whether it provides the right incentive structure for data product producers. Specifically, does it encourage the creation of more valuable data products while discouraging duplication and redundancy?

If the metric is tied to organizational reward systems—whether reputational or economic—it can influence producer behavior. Ideally, producers will be motivated to increase the relative value of their offerings compared to others in the ecosystem. But what strategies would this metric actually incentivize?

Under the proposed metric, producers are likely to pursue two main strategies to maximize the value of their offerings:
expanding their portfolio by creating more data products or focusing on uniqueness, ensuring their products are distinct and not overlapping with others.\par

To understand why, consider the probabilistic structure of consumer choices.

\subsubsection*{Probabilistic Model of Consumer Choice}

Let $m$ be the number of consumers $|C|=m$ and $n$ the number of offerings $|O|=n$. 

In this model, we assume consumer indifference among the available subsets of offerings—not as a reflection of actual behavior, but as a simplifying assumption in the absence of specific preference data. Accordingly, we adopt a uniform distribution over the power set of offerings, treating all possible combinations as equally probable to enable unbiased modeling of aggregate choice behavior.

Therefore, under this assumption, each consumer can choose any subset of offerings, resulting in $2^n$ possible choices per consumer. The total number of configurations across all consumers is:

\[
|S| = (2^n)^m
\]

The highest possible value $v(o_i)$ is $m$, and it is only achieved if all $m$ consumers choose $o_i$ and only $o_i$. The probability that a single offering is chosen by all consumers—thus achieving the maximum possible value $v(o_i) = m$ —is:

\[
P(v(o_i)) = m = \frac{1}{(2^n)^m}
\]

This extremely low probability highlights how rare it is for a single offering to dominate, reinforcing the value of uniqueness.

\subsubsection*{Illustrative Example}

Given 2 offerings and 2 consumers, there are 16 possible configurations of $sb : C \rightarrow P(O) $. The highest $v(o_1)=2$, and the probability of $v(o_1)=2$ is $\frac{1}{16}$.

\begin{center}
\setlength{\tabcolsep}{12pt}
\renewcommand{\arraystretch}{1.3}
\begin{tabular}{ |c|c|c|c| } \hline
$sb(c_1)$ & $sb(c_2)$ & $v(o_1)$ & $v(o_1)$ \\ \hline
$\{o_1\}$ & $\{o_1\}$ & $\frac{1}{1} + \frac{1}{1}$ & 2\\ \hline
$\{o_1\}$ & $\{o_1, o_2\}$ & $\frac{1}{1} + \frac{1}{2}$ & 1$\frac{1}{2}$\\ \hline
$\{o_1\}$ & $\{o_2\}$ & $\frac{1}{1}$ & 1\\ \hline
$\{o_1\}$ & $\{\emptyset\}$ & $\frac{1}{1}$ & 1\\ \hline

$\{o_1, o_2\}$ & $\{o_1\}$ & $\frac{1}{2} + \frac{1}{1}$ & 1$\frac{1}{2}$\\ \hline 
$\{o_1, o_2\}$ & $\{o_1, o_2\}$ & $\frac{1}{2} + \frac{1}{2}$ & 1\\ \hline 
$\{o_1, o_2\}$ & $\{o_2\}$ & $\frac{1}{2}$ & $\frac{1}{2}$\\ \hline 
$\{o_1, o_2\}$ & $\{\emptyset\}$ & $\frac{1}{2}$ & $\frac{1}{2}$\\ \hline

$\{o_2\}$ & $\{o_1\}$ & $\frac{1}{1}$ & 1\\ \hline
$\{o_2\}$ & $\{o_1, o_2\}$ & $\frac{1}{2}$ & $\frac{1}{2}$\\ \hline
$\{o_2\}$ & $\{o_2\}$ & $0$ & 0\\ \hline
$\{o_2\}$ & $\{\emptyset\}$ & $0$ & 0\\ \hline

$\{\emptyset\}$ & $\{o_1\}$ & $\frac{1}{1}$ & 1\\ \hline
$\{\emptyset\}$ & $\{o_1, o_2\}$ & $\frac{1}{2}$ & $\frac{1}{2}$\\ \hline
$\{\emptyset\}$ & $\{o_2\}$ & $0$ & 0\\ \hline
$\{\emptyset\}$ & $\{\emptyset\}$ & $0$ & 0\\ \hline

\end{tabular}
\end{center}
This example shows that when \textbf{all consumers choose all offerings}, each offering receives equal value $v(o_i)=1$, discouraging undifferentiated products.\par
Interestingly, when \textbf{each consumer selects a different offering}, each product also receives a score of exactly 1. While this reflects perfect uniqueness, it also results in uniform valuation across all offerings—no product stands out. This reveals a subtle insight: the metric does not reward uniqueness alone, but rather distinctiveness that resonates with multiple consumers. To achieve higher value, a product must not only be different, but strategically differentiated—designed to attract exclusive attention from a meaningful subset of users.\par 
If \textbf{no consumer selects an offering}, its value is zero: \( v(o_i) = 0 \). While this is mathematically straightforward, it reinforces a core principle of the metric—in the absence of engagement, even a potentially useful product contributes nothing to the overall valuation. This highlights the model’s emphasis on realized utility over unrealized value.\par

\subsubsection*{Implications for Producer Strategy}
To illustrate how the metric guides producer behavior, consider two offerings: \( o_i \) and \( o_j \). The value of \( o_i \) increases relative to \( o_j \) when more consumers subscribe to \( o_i \) exclusively, rather than subscribing to both or only to \( o_j \). Formally, this condition is expressed as:

\[
\left|\left\{c_l \mid o_i \in sb(c_l) \land o_j \notin sb(c_l)\right\}\right| > \left|\left\{c_l \mid o_j \in sb(c_l) \land o_i \notin sb(c_l)\right\}\right|
\]

This compares two groups: consumers who uniquely prefer \( o_i \) over \( o_j \) and consumers who uniquely prefer \( o_j \) over \( o_i \).

When the first group is larger, \( o_i \) earns a higher score, reflecting its differentiated appeal. This dynamic discourages producers from creating redundant offerings that overlap with existing ones. Instead, it incentivizes them to design products that fulfill unmet needs or offer unique value propositions—ultimately contributing to a more diverse and efficient data product ecosystem.

\subsection{Evaluating the Metric’s Role in Modeling Gross Data Product Value}
In general economics, it is assumed that increased production and consumption contribute to greater economic wealth, commonly measured as \textbf{Gross Product Value}—the total market value of all final goods and services produced. Ideally, the proposed metric should play a similar role in data economics within a company, contributing to what we refer to as the \textbf{gross data product value}.

We define gross data product value as the sum of the values of all offerings:
\[
V = \sum v(o)
\]

We observe that this sum increases with the growth in both production and consumption of data products. For example:

\begin{itemize}
    \item If all $m$ consumers choose a single offering $o_i$, then $V = m$. Therefore if the number of consumers increases to $|C|=m+1$, then $V$ also increases $V = m + 1$.
    \item If $m > n$, and each of $m$ consumers chooses a different offering, then $V = n$. Therefore if the number of offerings increases to $|O|=n+1$, then $V$ also increases  $V = n + 1$.
    \item This is also the case when each of the m consumers chooses all offerings. In this scenario $V = n$, and if the number of offerings increases to $|O|=n+1$, then $V$ also increases to $V = n + 1$.  
\end{itemize}

To farther illustrate how $V$ changes based on the number of offerings and their consumption frequency, we present the following scenarios:

\begin{quote} \textbf{Fixed consumption:} Each offering is consumed by a fixed number of consumers (e.g., 5). As the number of offerings increases, the total value $V$ increases linearly.\par
\textbf{Increasing consumption:} Each successive offering is consumed by more consumers than the previous one (e.g., 1, 2, 3, ...). This results in a faster-than-linear growth in $V$.\par
\textbf{Random consumption:} Each offering is consumed by a random number of different consumers. The total value fluctuates but generally increases with more offerings.\par
\end{quote}

These examples demonstrate that the metric responds to both the \textbf{quantity of offerings} and the \textbf{frequency of their consumption}, supporting the idea that it captures growth in data economics.

\begin{figure}[htpb]
    \centering
    \includegraphics[width=0.85\textwidth]{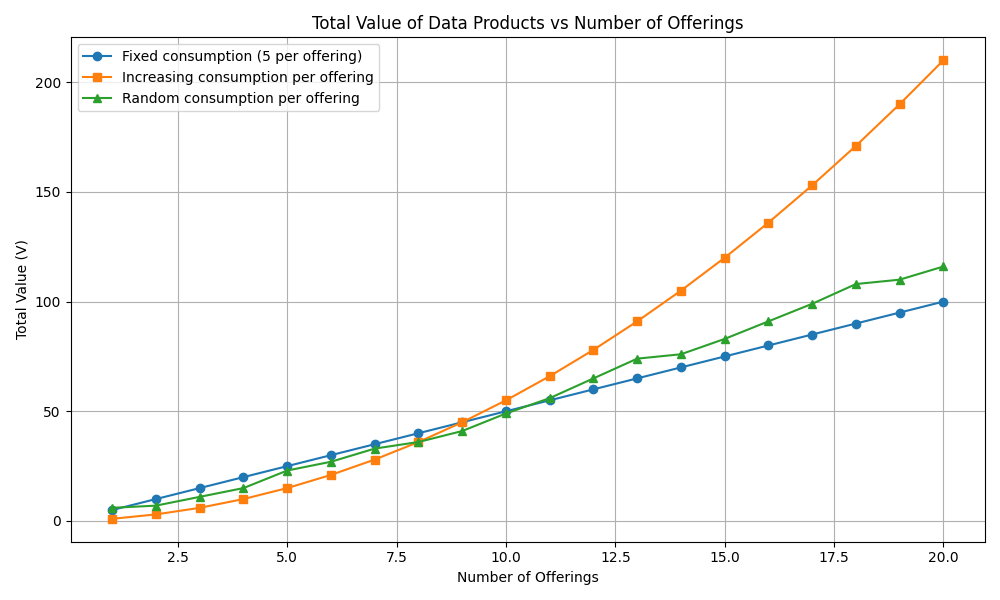}
    \caption{Total value of data products ($V$) as a function of the number of offerings and consumption frequency.}
    \label{fig:data_value_growth}
\end{figure}

As shown in Figure~\ref{fig:data_value_growth}, the metric reflects intuitive behavior: more offerings and higher consumption lead to greater total value. This supports the analogy to economic growth and reinforces the usefulness of the metric in evaluating data product performance.

Therefore, the GDaP formula will initially reward both the quantity of offerings and their consumption frequency to maximize total value. 

\begin{figure}[htpb]
    \centering
    \includegraphics[width=0.3\linewidth]{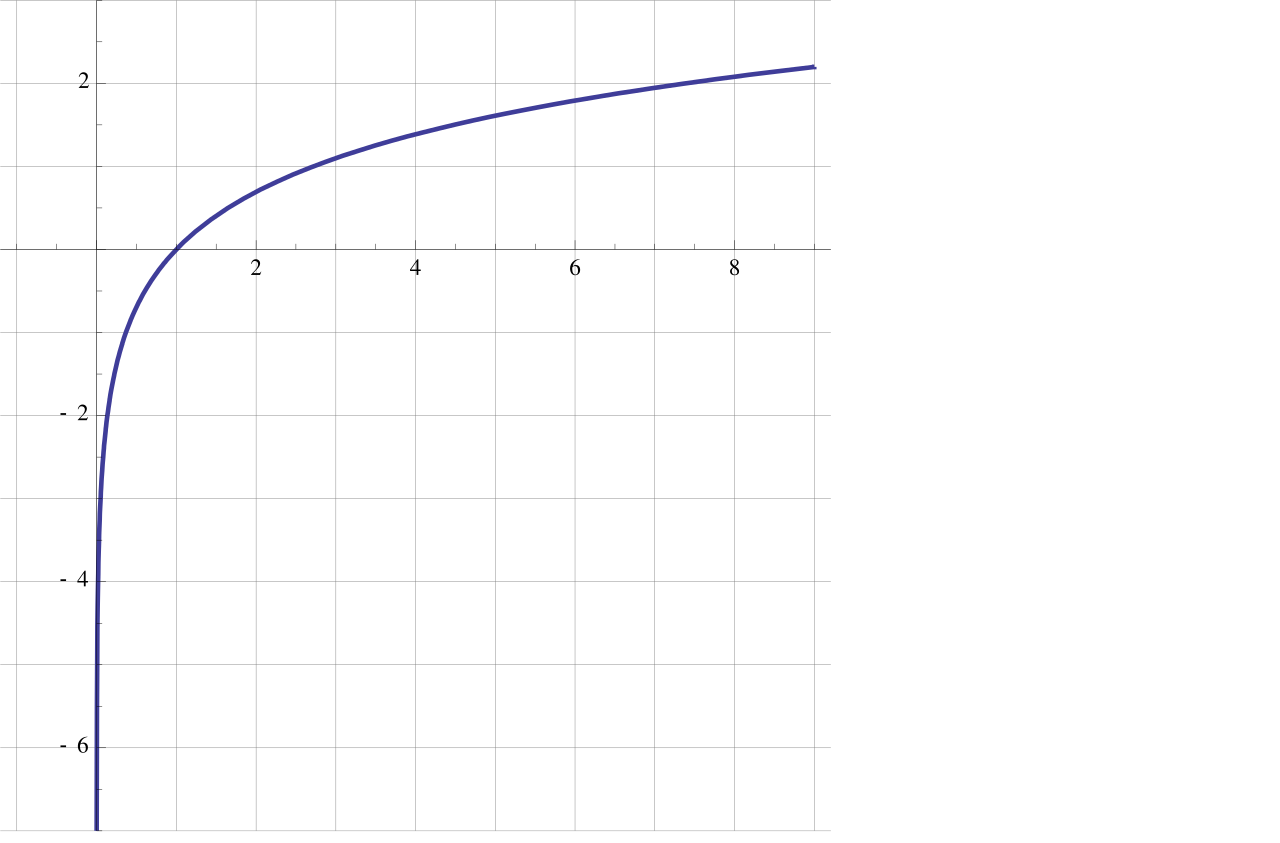}
    \caption{\label{log_growth}Logarithmic growth. Source: Wikipedia}
\end{figure}

Over time, however, as consumer attention and organizational capacity become saturated, each additional product yields diminishing marginal value. This saturation effect naturally slows the rate of new product introduction, shifting focus toward optimizing and deepening engagement with existing offerings. As a result, we expect the GDaP to follow \textbf{a logarithmic growth} pattern, consistent with general economic principles where diminishing returns emerge in mature markets \cite{wikipedia_market_saturation}. This reflects a transition from expansion-driven growth to value-driven refinement, characteristic of efficient and sustainable data economies (see Figure~\ref{log_growth}).

Importantly, logarithmic growth does not imply a decline in the number of data products. Rather, as efficiency and automation improve, data economics will continue to generate value—freeing up consumer attention and organizational capacity. This creates new opportunities for introducing offerings, even within a saturated environment.

\section{Conclusion and Future Considerations}
This paper introduced a novel, attention-based metric for valuing data products in non-monetary exchanges. Unlike traditional or popularity-based approaches, it quantifies value solely through consumer choice behavior, enabling fairer and more strategic evaluation of data offerings.

A key innovation lies in the metric’s ability to infer value without requiring external market signals or experimental data. By embedding attention directly into the valuation formula, it offers a scalable and intuitive framework for internal data product ecosystems, where monetary pricing is absent but strategic decision-making still demands rigorous valuation.

The metric’s formulation within a cooperative game-theoretic framework further provides a principled basis for fair value allocation.

Further analysis reveals a structural pattern in data consumption: the presence of \textbf{base products}, which are essential to a broad range of consumers and thus highly popular, and \textbf{complementary products}, which serve more specific needs and are consequently less popular. In many scenarios, consumers derive value from a combination of both.

For example, a coffee enthusiast may require an espresso machine (base product) and a bag of coffee beans (complementary product) to prepare espresso. Similarly, a business analyst might need access to a ``Customer 360'' dataset (base product) and a ``Customer Lifetime Value'' report (complementary product) to inform strategic decisions.

Empirical research supports this pattern: most digital marketplace consumers select from both base (``big head'') and complementary (``long tail'') products~\cite{goel2009anatomy}.

\medskip

Building on this insight, future research should explore value-multiplying bundling strategies, examining how combinations of base and complementary products create synergies that increase overall value, and develop methods for quantifying the additional value derived from such combinations. Game-theoretic approaches—particularly the Shapley value—offer a natural framework for attributing value among interdependent products and for modeling complementarity and bundling effects in data product ecosystems.



\bibliographystyle{alpha}
\bibliography{main}

\end{document}